\documentclass[twocolumn]{jpsj3} 

\renewcommand{\refname}{}

\usepackage{bm}

\title{High-$T_c$ Superconductivity with $T_c$ = 52 K under Antiferromagnetic Order in Five-layered Cuprate Ba$_2$Ca$_4$Cu$_5$O$_{10}$(F,O)$_2$ with $T_N$ = 175 K: $^{19}$F- and Cu-NMR Studies}

\author{Sunao \textsc{Shimizu}\thanks{E-mail: shimizu@nmr.mp.es.osaka-u.ac.jp}, 
Shin-ichiro \textsc{Tabata},
Hidekazu \textsc{Mukuda}, 
Yoshio \textsc{Kitaoka}, \\
Parasharam M. \textsc{Shirage}$^{1}$, 
Hijiri \textsc{Kito}$^{1}$,
and Akira \textsc{Iyo}$^{1}$}

\inst{Graduate School of Engineering Science, Osaka University, Toyonaka, Osaka 560-8531, Japan \\
$^{1}$National Institute of Advanced Industrial Science and Technology (AIST), Umezono, Tsukuba 305-8568, Japan \\}

\abst{We report on the observation of high-$T_c$ superconductivity (SC) emerging with the background of an antiferromagnetic (AFM) order in the five-layered cuprate Ba$_2$Ca$_4$Cu$_5$O$_{10}$(F,O)$_2$ through $^{19}$F-NMR and zero-field Cu-NMR studies. The measurements of spectrum and nuclear spin-lattice relaxation rates $^{19}(1/T_1)$ of $^{19}$F-NMR give convincing evidence for the AFM order taking place below $T_N$ = 175 K and for the onset of SC below $T_c$ = 52 K, hence both coexisting. 
The zero-field Cu-NMR study has revealed that AFM moments at Cu sites are 0.14 $\mu_B$ at outer CuO$_2$ layers and 0.20 $\mu_B$ at inner ones. We remark that an intimate coupling exists between the AFM state and the SC order parameter below $T_c$ = 52 K; the spin alignment in the AFM state is presumably changed in the SC-AFM mixed state.}

\kword{high-$T_c$ superconductivity, copper-oxide, antiferromagnetism, NMR, multilayer}

\begin{document}
\maketitle


Since the discovery of high-$T_c$ superconductivity (HTSC), one of the most challenging issues in condensed matter physics has been to clarify the mechanism by which it is induced. In the phase diagrams of HTSC with doping, a long-standing problem is the interplay between HTSC and antiferromagnetism (AFM) from both experimental~\cite{Weidinger,Tranquada,Niedermayer,Lee,Sidis,Lake,Sanna,Miller,Das,Stock,Haug,Coneri} and theoretical~\cite{Chen,Giamarchi,Inaba,Anderson1,Zhang,Himeda,Kotliar,TKLee,Demler,Shih,Yamase,Paramekanti,Anderson2,Senechal,Capone,Pathak} points of view.  
In extensive researches on underdoped regimes, the persistence of magnetic moments has been reported even in the SC phases in La$_{2-x}$Sr$_x$CuO$_4$ (LSCO) and YBa$_2$Cu$_3$O$_{6+y}$ (YBCO). 
In fact, Sidis {\it et al.} reported that a commensurate AFM order in YBCO$_{6.5}$ with $T_c$ = 55 K may take place on a nanosecond time scale~\cite{Sidis}.
On the other hand, in multilayered cuprates such as five-layered compounds HgBa$_2$Ca$_4$Cu$_5$O$_{12+\delta}$~(Hg-1245)~ \cite{Mukuda} and four-layered ones Ba$_2$Ca$_3$Cu$_4$O$_{8}$(F,O)$_2$~\cite{ShimizuJPSJ}, we have demonstrated  that HTSC uniformly coexists with a completely static AFM order on a single CuO$_2$ plane. 
It is important to address characteristics of AFM order below the ordering temperature $T_N$ and below $T_c$ in order to highlight how the onset of SC affects an AFM state.  

In this letter, we report $^{19}$F-NMR and zero-field Cu-NMR studies on an underdoped five-layered compound Ba$_2$Ca$_4$Cu$_5$O$_{10}$(F,O)$_2$ with $T_c$ = 52 K. The present NMR studies reveal that AFM order takes place below $T_N$ = 175 K with AFM moments $M_{AFM}$(OP) $\sim$ 0.14 $\mu_B$ at OP and $M_{AFM}$(IP) $\sim$ 0.20 $\mu_B$ at IP. We demonstrate that the onset of the SC order parameter below $T_c$ = 52 K brings about an intimate coupling with the AFM moments. 

\begin{figure}[htpb]
\begin{center}
\includegraphics[width=1.0\linewidth]{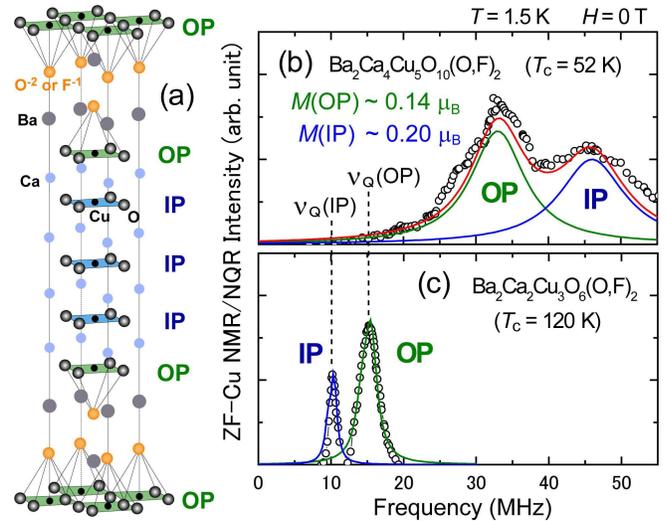}
\end{center}
\caption{\footnotesize (color online) (a) Crystal structure of Ba$_2$Ca$_4$Cu$_5$O$_{10}$(F,O)$_2$. There are two kinds of CuO$_2$ layers: outer planes (OPs) and inner ones (IPs). (b) Zero-field Cu-NMR spectra at $T$ = 1.5 K. In the spectra, $^{63}$Cu and $^{65}$Cu are not resolved due to the poor frequency resolution that is related to a week signal-noise ratio. Therefore, each spectrum for OP and IP is temtatively represented as a single Lorentzian. (c) NQR spectra for three-layered compound Ba$_2$Ca$_2$Cu$_3$O$_6$(F,O)$_2$ with $T_c$ = 120 K. Dashed lines denote NQR frequencies for OP and IP.}
\label{fig:ZF}
\end{figure}

The polycrystalline powder sample of five-layered Ba$_2$Ca$_4$Cu$_5$O$_{10}$(F,O)$_2$~(0245F) was prepared by a  high-pressure synthesis technique~\cite{Iyo1}. The nominal composition of the sample used in this study is nearly Ba$_2$Ca$_4$Cu$_5$O$_{10}$F$_2$, although it is difficult to precisely determine the actual fraction of F$^{1-}$ and O$^{2-}$ at apical sites \cite{Shirage,ShimizuPRB}. Figure \ref{fig:ZF}(a) illustrates the crystal structure of 0245F with two kinds of CuO$_2$ planes: outer planes (OPs) and inner ones (IPs). The superconducting transition temperature $T_c$ = 52 K was uniquely determined by the onset of SC diamagnetism measured by a dc SQUID magnetometer. In the multilayered systems with apical-fluorine (F), an increase in the nominal fraction of oxygen O$^{2-}$ at apical-F sites increases the hole doping level and hence $T_c$~\cite{Iyo1}. The respective hole doping levels  at OP and IP are estimated to be $p$(OP) $\sim$~0.06-0.07 and $p$(IP) $\sim$~0.04-0.05 from the measurement of the spin part of the Knight shift at $T$ = 300 K. Powder X-ray diffraction analysis shows that the compound is comprised of almost a single phase. For NMR measurements, the powder sample was aligned along the $c$-axis in an external field $H_0$ of 16 T and fixed using stycast 1266 epoxy. The F-NMR experiments were performed by a conventional spin-echo method in the temperature ($T$) range from 1.5 to 300 K with $H_0$ parallel to the $c$-axis, and the Cu-NMR experiments were at zero-field.


In general, the Hamiltonian for Cu nuclear spin ($I=3/2$) with an axial symmetry is described in terms of the Zeeman interaction ${\cal H}_{Z}$ due to a magnetic field $H$, and the nuclear-quadrupole interaction ${\cal H}_{Q}$ as follows:
\begin{eqnarray}
{\cal H}&=&{\cal H}_Z+{\cal H}_Q  \notag \\
        &=&-\gamma_N \hbar {\bm I} \cdot {\bm H}+\frac{e^{2}qQ}{4I(2I-1)}(3I_{z^{\prime}}^2-I(I+1)),
\label{eq:hamiltonian}
\end{eqnarray}
where $\gamma_{N}$ is the Cu nuclear gyromagnetic ratio, $eQ$ the nuclear quadrupole moment, and $eq$ the electric field gradient at a Cu site. In ${\cal H}_{Q}$, the nuclear quadrupole resonance (NQR) frequency is defined as $\nu_{Q}=e^{2}qQ/2h$.
In non-magnetic substances, an NQR spectrum is observed due to the second term in Eq.~(\ref{eq:hamiltonian}) when $H$ = $H_0$ = 0. 
On the other hand, in magnetically ordered substances, internal magnetic fields $H_{int}$ is induced at Cu sites; in addition to the second term, the first term in Eq.~(\ref{eq:hamiltonian}) contributes to the nuclear hamiltonian even if $H_0$ = 0 T. 
Therefore, the Cu-NQR and the Cu-NMR at $H_0$ = 0 ($H$ = $H_{int}$) can sensitively detect the onset of a magnetically ordered state.

Figure \ref{fig:ZF}(c) presents the Cu-NQR spectrum of a three-layered compound Ba$_2$Ca$_2$Cu$_3$O$_6$(F,O)$_2$ (0223F) with $T_c$ = 120 K. The respective NQR frequencies are $^{63}\nu_Q$(IP) = 10.7 MHz at IP and $^{63}\nu_Q$(OP) = 15.5 MHz at OP, which coincide with values observed in most multilayered cuprates~\cite{Zheng,Magishi,ShimizuJPSJ}. This result assures that the compound is a paramagnetic superconductor. By contrast, a zero-field Cu-NMR spectrum of 0245F at $T$ = 1.5 K in Fig.~\ref{fig:ZF}(b) shows two peaks at around 30 MHz and 45 MHz, which are significantly larger than the NQR frequencies in Fig.~\ref{fig:ZF}(c). According to Eq.~(\ref{eq:hamiltonian}), resonance frequencies increase when $H_{int}$ exists  in association with an onset of AFM order. The respective spectra around 30 MHz and 45 MHz are assigned to OP and IP, indicating that $H_{int}$(OP) is smaller than $H_{int}$(IP). This is because an AFM moment in proportion to $H_{int}$ is small with doping. The obtained values $H_{int}$(OP) $\sim$ 2.8 T and $H_{int}$(IP) $\sim$ 4.1 T allow us to estimate $M_{AFM}$(OP) $\sim$ 0.14 $\mu_{B}$ and $M_{AFM}$(IP) $\sim$ 0.20 $\mu_{B}$, respectively. Here, we use the relation of $H_{int}$ = $|A_{hf}|M_{AFM}$ = $|A-4B|M_{AFM}$, where $A$ and $B$ are the on-site hyperfine field and the supertransferred hyperfine field, respectively. $A$ $\sim$ 3.7 T/$\mu_{B}$ and $B$(IP) $\sim$ 6.1 T/$\mu_{B}$ are assumed, which is typical values of multilayered cuprates in underdoped regions \cite{Kotegawa2004,ShimizuP}. Here, note that there is no phase separation between magnetic phases and SC phases in the present sample because no NQR signal pointing to the presence of paramagnetic SC phases is observed around frequencies marked by the dashed lines in Fig.~\ref{fig:ZF}(b).  

\begin{figure}[tpb]
\begin{center}
\includegraphics[width=0.74\linewidth]{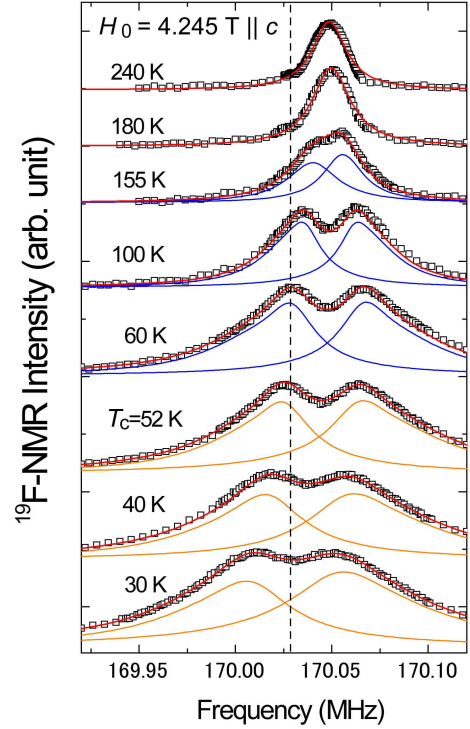}
\end{center}
\caption{\footnotesize (color online)  $^{19}$F-NMR spectra for Ba$_2$Ca$_4$Cu$_5$O$_{10}$(F,O)$_2$ at various temperatures with $H_0$ parallel to the $c$-axis. The solid lines are spectral simulations. The dashed line indicates $^{19}K_c$ = 0.}
\label{fig:F-spe}
\end{figure}

It is difficult to deduce the $T$ dependences of $M_{AFM}$(OP) and $M_{AFM}$(IP) in the AFM state from the Cu-NQR or the zero-field Cu-NMR measurements because of the extremely short nuclear spin relaxation time at Cu sites. Instead, $^{19}$F-NMR is measured to probe the $T$ dependence of the internal field ${\bm H_{int}}$(F) at apical-F sites, which is induced by either $M_{AFM}$(OP) or 
$M_{AFM}$(IP). Figure \ref{fig:F-spe} shows the $T$ dependence of $^{19}$F-NMR spectra obtained by sweeping the frequencies at $H_0$ = 4.245 T parallel to the $c$-axis. A sharp spectrum is observed with a single peak at $T$ = 240 K, but the spectrum splits into two peaks below $T$ = 175 K. 

\begin{figure}[tpb]
\begin{center}
\includegraphics[width=0.93\linewidth]{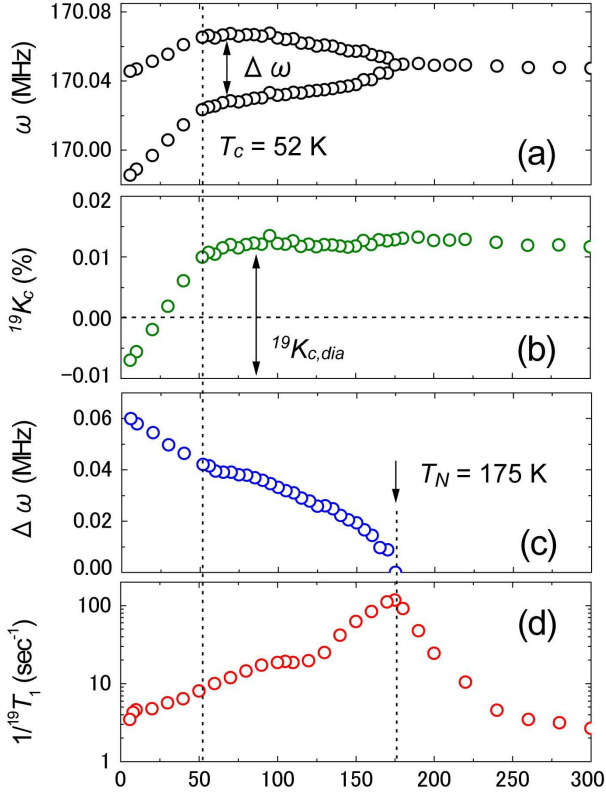}
\end{center}
\caption{\footnotesize (color online) Temperature dependences of (a) the resonance frequency $\omega$ of $^{19}$F-NMR and of (b) the Knight shift $^{19}K_c$. The decrease in $^{19}K_c$ below $T_c$ = 52 K is mainly associated with a SC diamagnetic shift. (c) Temperature dependence of splitting of two peaks $\Delta\omega$ in $^{19}$F-NMR spectra. The spectra split into two below $T_N$ = 175 K due to the $H_{\rm int}$ at apical-F sites. (d) Temperature dependence of nuclear spin-lattice relaxation rate 1/$^{19}T_1$. The peak in 1/$^{19}T_1$ at $T_N$ = 175 K is associated with the critical slowing down of spin fluctuations at the IP and the OP.}
\label{fig:Fdata}
\end{figure}

Figure \ref{fig:Fdata}(a) presents the $T$ dependence of resonance frequency $\omega$ in $^{19}$F-NMR spectra, which is deduced through spectral simulations shown by solid lines in Fig.~\ref{fig:F-spe}.  
The resonance frequency of $^{19}$F-NMR at $H_0$ parallel to the $c$-axis is expressed by  
\begin{eqnarray}
\omega \simeq {^{19}\gamma_N} H_0~(1+{^{19}K_c})\pm {^{19}\gamma_N}|H_{int,c}({\rm F})|,
\label{eq:Hint}
\end{eqnarray}
where $H_{int,c}$(F) is the component of ${\bm H_{int}}$(F) along the $c$-axis, $^{19}\gamma_{N}$ 
the $^{19}$F nuclear gyromagnetic ratio, and $^{19}K_c$ the Knight shift.  The plus (minus) sign of $|H_{int,c}({\rm F})|$ in Eq.~(\ref{eq:Hint}) corresponds to its parallel (antiparallel) component along the $c$-axis and hence $\Delta\omega$ in Fig.~\ref{fig:Fdata}(c) is proportional to 2$\times^{19}\gamma_N |H_{int,c}({\rm F})|$.
Upon cooling, its splitting $\Delta\omega$, shown in Fig.~\ref{fig:Fdata}(c), increases due to the development of ${\bm H_{int}}$(F) in proportion to $M_{AFM}$(OP,IP) below the N\'eel temperature $T_N$ = 175 K. 
The onset of the AFM order at $T_N$ = 175 K is also corroborated by the peak in the nuclear spin-lattice relaxation rate 1/$^{19}T_1$ as shown in Fig.~\ref{fig:Fdata}(d). 

\begin{figure}[tpb]
\begin{center}
\includegraphics[width=1\linewidth]{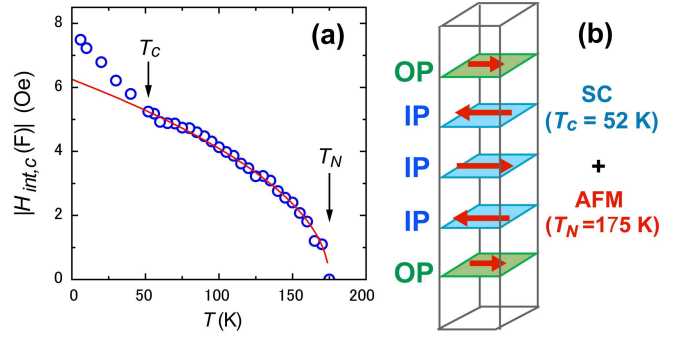}
\end{center}
\caption{\footnotesize (color online)  (a) Temperature dependence of $|H_{int,c}({\rm F})|$ estimated from $\Delta \omega$ = 2$\times^{19}\gamma_N |H_{int,c}({\rm F})|$. Note that $M_{AFM}$(OP) is proportional to $|H_{int,c}({\rm F})|$ (see text). The solid line is $H_{int,c}$(F)$\propto M_{AFM}=M_{AFM}(0)(1-T/T_N)^{0.5}$, which is in good agreement with the experiment down to $T_c$ = 52 K. (b) Illustration of ground state property of Ba$_2$Ca$_4$Cu$_5$O$_{10}$(F,O)$_2$ with $T_c$ = 52 K and $T_N$ = 175 K.}
\label{fig:M}
\end{figure}

Next, we deal with the $T$ dependence of ${\bm H_{int}}$(F) shown in Fig.~\ref{fig:M}(a). 
If a direction of $M_{AFM}$  were in the basal planes with a wave vector $Q$ = ($\pi$,$\pi$), as denoted in Fig.~\ref{fig:M}(b), $H_{int,c}$(F) could be cancelled out at the apical-F site; the F ions are located at a magnetically symmetric position. Therefore, the presence of $H_{int,c}$(F) points to the fact that $M_{AFM}$ is presumably directed out of the planes. This out-of-plane  canting of the spin direction, which has been reported for La$_2$CuO$_4$ \cite{Thio,Coffey,Tsuda,Kastner} or bi-layered TlBa$_2$YCu$_2$O$_7$~\cite{Goto}, was interpreted in terms of the Dzyaloshinsky-Moriya (DM) type of anisotropic exchange interaction. Generally, the DM interaction is induced by the spin-orbit interaction, when a middle point between two spins is not an inversion center~\cite{Dzyaloshinsky,Moriya}. 
In the case of 0245F, the Cu site at OP is located with the out-of-plane distance of $\sim$ 0.015 ${\rm \AA}$, which breaks the inversion symmetry at the middle point of the two spins on OP. Accordingly, $H_{int,c}$(F) is induced by the out-of-plane canting of $M_{AFM}$(OP) due to the DM interaction. On the other hand, $M_{AFM}$(IP) is expected to be in the planes because of the square oxygen coordination without apical site.
As a result, $H_{int,c}$(F) is concluded to be produced only by $M_{AFM}$(OP), not by $M_{AFM}$(IP) 

As for the origin of the internal magnetic field at apical sites, it was reported that the dominant contribution is not a classical dipole field but a transferred hyperfine field from the nearest neighboring Cu spins~\cite{Kambe}.  When using the transferred hyperfine coupling constant between the plane Cu spin and the apical oxygen nuclear in Tl$_2$Ba$_2$CuO$_y$~\cite{Kambe}, the canting angle is estimated to be as small as a few degrees.
In order to shed light on the $T$ evolution of $M_{AFM}(T)$, the solid line in Fig.~\ref{fig:M}(a) displays a variation of $H_{int,c}$(F) $\propto$ $M_{AFM}(T)$ = $M_{AFM}(0)(1-T/T_N)^{0.5}$, which is in good agreement with the experiment down to $T_c$ = 52 K. 
This power-law variation of $M_{AFM}(T)$ in the AFM state down to $T_c$ = 52 K coincides with those for slightly-doped LSCO compounds that exhibit AFM ground states~\cite{Borsa}.
Here, we note that $H_{int,c}$(F) shows an additional increase as $T$ falls below $T_c$; the increase in $H_{int,c}$(F) below $T_c$ convinces one that the onset of a SC order parameter is actually coupled with $M_{AFM}$(OP) in the SC-AFM coexisting state. 

Finally, we deal with the SC properties in 0245F. The $T$ dependence of the Knight shift $^{19}K_c$ is displayed in Fig.~\ref{fig:Fdata}(b), where $^{19}K_c$ below $T_N$ = 175 K is estimated as the average value of the two peaks, being $T$ independent down to $T_c$ = 52 K. Usually in cuprates, the spin-components of the Knight shift start to decrease upon cooling far from above $T_c$, which is due to the opening of pseudogaps \cite{REbook}. The reason that $^{19}K_c$ is $T$-independent above $T_c$ is that the spin component in $^{19}K_c$ is small owing to the small hyperfine coupling between $^{19}$F and Cu-3d spins as reported 
in the literature~\cite{Kambe}. By contrast, $^{19}K_c$ markedly decreases due to the appearance of SC diamagnetism below $T_c$ = 52 K. The reduction of $^{19}K_c$, which is in association with the onset of HTSC, takes place under the background of the AFM order, providing firm evidence for the uniform coexisting state of AFM and SC at a microscopic level. It is likely that the onset of HTSC with a d-wave symmetry with spin-singlet pairing decreases the size of $M_{AFM}$(OP) due to the formation of coherent spin-singlet states over the sample, so that the additional increase in $H_{int,c}$(F) below $T_c$ may not be due to an increase of $M_{AFM}$(OP) but an increase of the out-of-plane canting angle in the SC mixed state. In any case, this finding is the first in the HTSC phenomena to demonstrate an intimate coupling between the SC order parameter and $M_{AFM}$.

In summary, we have reported $^{19}$F-NMR and zero-field Cu-NMR measurements on the underdoped five-layered cuprate Ba$_2$Ca$_4$Cu$_5$O$_{10}$(F,O)$_2$ with $T_c$ = 52 K. The ground-state property is schematically shown in Fig.~\ref{fig:M}(b).
The present NMR studies have provided firm evidence for the AFM order taking place below $T_N$ = 175 K with respective AFM moments of 0.14 and 0.20 $\mu_B$ at outer and inner CuO$_2$ layers, showing the uniform mixing with the SC order parameter below $T_c$ = 52 K. We have highlighted the fact that an intimate coupling emerges between the AFM state and the SC order parameter below $T_c$ = 52 K. The onset of SC order parameter leads to an evolution of the spin alignment in the AFM-SC mixed state, resulting in the increase of the internal field along the c-axis at apical F sites. Further detailed experiments on single crystals are required to clarify the significant  change in a spin structure in going from an AFM state to the SC-AFM mixed state.

The authors are grateful to M. Mori for his helpful discussions. This work was supported by Grant-in-Aid for Specially Promoted Research (20001004) and by the Global COE Program (Core Research and Engineering of Advanced Materials-Interdisciplinary Education Center for Materials Science) from the Ministry of Education, Culture, Sports, Science and Technology (MEXT), Japan.



\begin{thebibliography}{99} 


\bibitem{Weidinger} A. Weidinger, Ch. Niedermayer, A. Golnik, R. Simon, and E. Recknagel: Phys. Rev. Lett. {\bf 62} (1989) 102. 
\bibitem{Tranquada} J. M. Tranquada, J. D. Axe, N. Ichikawa, A. R. Moodenbaugh, Y. Nakamura, and S. Uchida: Phys. Rev. Lett. {\bf 78} (1997) 338. 
\bibitem{Niedermayer} Ch. Niedermayer, C. Bernhard, T. Blasius, A. Golnik, A. Moodenbaugh, and J. I. Budnick: Phys. Rev. Lett. {\bf 80}3 (1998) 384. 
\bibitem{Lee}  Y. S. Lee, R. J. Birgeneau, M. A. Kastner, Y. Endoh, S. Wakimoto, K. Yamada, R. W. Erwin, S.-H. Lee, and G. Shirane: Phys. Rev. B {\bf 60} (1999) 3643. 
\bibitem{Sidis} Y. Sidis, C. Ulrich, P. Bourges, C. Bernhard, C. Niedermayer, L. P. Regnault, N. H. Anderson, and B. Keimer: Phys. Rev. Lett. {\bf 86} (2001) 4100. 
\bibitem{Lake} B. Lake, H. M. Ronnow, N. B. Christensen, G. Aeppli, K. Lefmann, D. F. McMorrow, P. Vorderwisch, P. Smeibidl, N. Mangkorntong, T. Sasagawa, M. Nohara, H. Takagi, and T. E. Mason: Nature {\bf 415} (2002) 299. 
\bibitem{Sanna} S. Sanna, G. Allodi, G. Concas, A. D. Hillier, and R. DeRenzi: Phys. Rev. Lett. {\bf 93} (2004) 207001. 
\bibitem{Miller}  R. I. Miller, R. F. Kiefl, J. H. Brewer, F. D. Callaghan, J. E. Sonier, R. Liang, D. A. Bonn, and W. Hardy: Phys. Rev. B {\bf 73} (2006) 144509. 
\bibitem{Das} T. Das, R. S. Markiewicz, and A. Bansil: Phys. Rev. Lett. {\bf 98} (2007) 197004. 
\bibitem{Stock} C. Stock, W. J. L. Buyers, Z. Yamani, Z. Tun, R. J. Birgeneau, R. Liang, D. Bonn, and W. N. Hardy: Phys. Rev. B {\bf 77} (2008) 104513. 
\bibitem{Haug} D. Haug, V. Hinkov, A. Suchaneck, D. S. Inosov, N. B. Christensen, Ch. Niedermayer, P. Bourges, Y. Sidis, J. T. Park, A. Ivanov, C. T. Lin, J. Mesot, and B. Keimer: Phys. Rev. Lett. {\bf 103} (2009) 017001. 
\bibitem{Coneri} F. Coneri, S. Sanna, K. Zheng, J. Lord, and R. DeRenzi: Phys. Rev. B {\bf 81} (2010) 104507. 
\bibitem{Chen}  G. J. Chen, R. Joynt, F. C. Zhang, and C. Gros: Phys. Rev. B {\bf 42} (1990) 2662.
\bibitem{Giamarchi} T. Giamarchi and C. Lhuillier: Phys. Rev. B {\bf 43} (1991) 12943.
\bibitem{Inaba} M. Inaba, H. Matsukawa, M. Saitoh, and H. Fukuyama: Physica C {\bf 257} (1996) 299.
\bibitem{Anderson1} P.W. Anderson: {\it The Theory of Superconductivity in the High-$T_c$ Cuprate Superconductors} (Princeton Univ. Press, Princeton, 1997).
\bibitem{Zhang} S. C. Zhang: Science {\bf 275} (1997) 1089.
\bibitem{Himeda} A. Himeda and M. Ogata: Phys. Rev. B {\bf 60} (1999) R9935.
\bibitem{Kotliar} G. Kotliar, S. Y. Savrasov, G. Palsson, and G. Biroli: Phys. Rev. Lett. {\bf 87} (2001) 186401.
\bibitem{TKLee} T. K. Lee, C.-M. Ho, and N. Nagaosa: Phys. Rev. Lett. {\bf 90} (2003) 067001.
\bibitem{Demler} E. Demler, W. Hanke, and S. C. Zhang: Rev. Mod. Phys. {\bf 76} (2004) 909.
\bibitem{Shih} C. T. Shih, T. K. Lee, R. Eder, C. Y. Mou, and Y. C. Chen: Phys. Rev. Lett {\bf 92} (2004) 227002.
\bibitem{Yamase} H. Yamase and H. Kohno: Phys. Rev. B {\bf 69} (2004) 104526.
\bibitem{Paramekanti} A. Paramekanti, M. Randeria and N. Trivedi: Phys. Rev. B {\bf 70} (2004) 054504.
\bibitem{Anderson2} P.W. Anderson, P.A. Lee, M. Randeria, T.M. Rice, N. Trivedi, and F.C. Zhang: J. Phys. Condens. Matter {\bf 16} (2004) R755.
\bibitem{Senechal} D. S\'en\'echal, P. L. Lavertu, M. A. Marois, and A. M. S. Tremblay: Phys. Rev. Lett. {\bf 94} (2005) 156404.
\bibitem{Capone} M. Capone and G. Kotliar: Phys. Rev. B {\bf 74} (2006) 054513.
\bibitem{Pathak} S. Pathak, V. B. Shenoy, M. Randeria, and N. Trivedi: Phys. Rev. Lett. {\bf 102} (2009) 027002.

\bibitem{Mukuda} H. Mukuda, Y. Yamaguchi, S. Shimizu, Y. Kitaoka, P. Shirage, and A. Iyo: J. Phys. Soc. Jpn. {\bf 77} (2008) 124706.
\bibitem{ShimizuJPSJ} S. Shimizu, H. Mukuda, Y. Kitaoka, H. Kito, Y. Kodama, P. M. Shirage, and A. Iyo: J. Phys. Soc. Jpn. {\bf 78} (2009) 064705.
\bibitem{Iyo1} A. Iyo, Y. Tanaka, M. Tokumoto, and H. Ihara: Physica C {\bf 366} (2001) 43. 
\bibitem{Shirage} P. M. Shirage, D. D. Shivagan, Y. Tanaka, Y. Kodama, H. Kito, and A. Iyo: Appl. Phys. Lett. {\bf 92} (2008) 222501.
\bibitem{ShimizuPRB} S. Shimizu, T. Sakaguchi, H. Mukuda, Y. Kitaoka, P. M. Shirage, Y. Kodama, and A. Iyo: Phys. Rev. B {\bf 79} (2009) 064505.
\bibitem{Zheng} G.-q. Zheng, Y. Kitaoka, K. Asayama, K. Hamada, H. Yamauchi, and S. Tanaka: Physica C {\bf 260} (1996) 197.
\bibitem{Magishi} K. Magishi, G.-q. Zheng, Y. Kitaoka, K. Asayama, K. Tokiwa, A. Iyo, and H. Ihara: Physica C {\bf 263} (1996) 375.
\bibitem{Kotegawa2004} H. Kotegawa, Y. Tokunaga, Y. Araki, G.-q. Zheng, Y. Kitaoka, K. Tokiwa, K. Ito, T. Watanabe, and A. Iyo: Phys. Rev. B {\bf 69} (2004) 014501.
\bibitem{ShimizuP} S. Shimizu, S.-i. Tabata, H. Mukuda, Y. Kitaoka, P. M. Shirage, H. Kito, and A. Iyo: submitted to Phys. Rev. B.
\bibitem{Tsuda} T. Tsuda, T. Shimizu, H. Yasuoka, K. Kishio, and K. Kitazawa: J. Phys. Soc. Jpn {\bf 57} (1988) 2908.
\bibitem{Kastner} M. A. Kastner, R. J. Birgeneau, T. R. Thurston, P. J. Picone, H. P. Jenssen, D. R. Gabbe, M. Sato, K. Fukuda, S. Shamoto, Y. Endoh, K. Yamada, and G. Shirane: Phys. Rev. B {\bf 38} (1988) 6636.
\bibitem{Thio} T. Thio, T. R. Thurston, N. W. Preyer, P. J. Picone, M. A. Kastner, H. P. Jenssen, D. R. Gabbe, C. Y. Chen, R. J. Birgeneau, and A. Aharony: Phys. Rev. B {\bf 38} (1988) 905.
\bibitem{Coffey} D. Coffey, T. M. Rice, and F. C. Zhang: Phys. Rev. B {\bf 44} (1991) 10112.
\bibitem{Goto} T. Goto, S. Nakajima, M. Kikuchi, Y. Syono, and T. Fukase: Phys. Rev. B {\bf 54} (1996) 3562.
\bibitem{Dzyaloshinsky} I. Dzyaloshinsky: Phys. Chem. Solids {\bf 4} (1958) 241.
\bibitem{Moriya} T. Moriya: Phys. Rev. {\bf 120} (1960) 91.
\bibitem{Kambe} S. Kambe, H. Yasuoka, A. Hayashi, and Y. Ueda: Phys. Rev. B {\bf 48} (1993) 6593.
\bibitem{Borsa} F. Borsa, P. Carretta, J. H. Cho, F. C. Chou, Q. Hu, D. C. Jhonston, A. Lascialfari, D. R. Torgeson, R. J. Gooding, N. M. Salem, and K. J. E. Vos: Phys. Rev. B {\bf 52} (1995) 7334.
\bibitem{REbook} R. E. Walstedt: {\it The NMR Probe of High-$T_c$ Materials} (Springer, Berlin, 2008).



\end{thebibliography}
\end{document}